\documentclass[nofootinbib,aps,onecolumn]{revtex4}
\usepackage{amsmath}
\usepackage{graphicx}
\usepackage{amssymb}
\usepackage{float}
\usepackage{mathpazo}
\usepackage{lineno}

\pdfoutput=1
\newcommand{\ignore}[1]{}

\newcommand{\be}{\begin{equation}}
\newcommand{\ee}{\end{equation}}
\def \ba#1\ea{\begin{align}#1\end{align}}
\newcommand{\bit}{\begin{itemize}}
\newcommand{\eit}{\end{itemize}}

\def \slashb#1{\setbox0=\hbox{$#1$}#1\hskip-\wd0\dimen0=5pt\advance
        \dimen0 by-\ht0\advance \dimen0 by\dp0\lower0.5\dimen0\hbox
          to\wd0{\hss \sl/\/ \hss}}

\input{epsf}
\input{tcilatex}
\begin{document}

\title{\textbf{Sin$^{2}\theta _{W}$ estimate and bounds on nonstandard
interactions at source and detector in the solar neutrino low-energy regime}}
\author{Amir N. Khan}
\email{ntrnphysics@gmail.com, khan8@mail.sysu.edu.cn}
\affiliation{School of Physics, Sun Yat-Sen University, Guangzhou, Guangdong, 510275, P.
R. China}
\author{Douglas W. McKay}
\email{dmckay@ku.edu}
\affiliation{Department of Physics and Astronomy, University of Kansas, Lawrence, KS
66045, USA}

\begin{abstract}
We explore the implications of the Borexino experiment's real time
measurements of the lowest energy part of the neutrino spectrum from the
primary \textit{pp} fusion process up to 0.420 MeV through the $^{7}$Be
decay at 0.862 MeV to the \textit{pep} reaction at 1.44 MeV. We exploit the
fact that at such low energies, the large mixing angle solution to the
Mikheyev-Smirnov-Wolfenstein matter effects in the sun are small for $^{7}$%
Be and $pep$ and negligible for $pp$. Consequently, the neutrinos produced
in the sun change their flavor almost entirely through vacuum oscillations
during propagation from the sun's surface and through possible nonstandard
interactions acting at the solar source and Borexino detector. We combine
the different NSI effects at source and detector in a single framework and
use the current Borexino data to bound NSI non-universal and flavor-
changing parameters at energies below the reach of reactor neutrino
experiments. We also study the implication of the current data for the weak-
mixing angle at this "low-energy frontier" data from the Borexino
experiment, where it is expected to be slightly larger than its value at the
Z mass. We find $\sin ^{2}\theta _{W}=0.224\pm 0.016$, the lowest
energy-scale estimate to date. Looking to the future, we use projected
sensitivities to solar neutrinos in next generation dedicated solar
experiments and direct dark matter detection experiments and find a
potential factor five improvement in determination of the weak-mixing angle
and up to an order of magnitude improvement in probing the NSI parameters
space.
\end{abstract}

\date{\today}
\pacs{xxxxx}
\maketitle

\section{ Introduction}

Experimental and theoretical studies of solar neutrinos have played a
special role in our understanding of solar structure and dynamics \cite{b&u1}
and the interaction and propagation of neutrinos themselves \cite{davis}.
Reviews of the current picture and some prospects for the future can be
found in Refs. \cite{M&S,ianni,haxton}. Here we focus on the lowest energy
range of solar neutrinos, whose sources are the \textit{pp}, $^{7}$Be and 
\textit{pep} processes. In recent years, the Borexino experiment \cite%
{Borexino_pp, Borexino_7Be, Borexino_pep, Borexino_review} has published for
the first time the discovery or strong evidence for direct detection of all
three of these neutrino sources, providing neutrino detection rates that can
probe the predictions of the value of the weak-mixing angle at and below an
MeV and the possibility of deviations from the predictions of the standard
model (SM) in the form of nonstandard interactions (NSIs).

Neutrinos produced at the Sun are detected at the Borexino detector using
the purely leptonic elastic processes e+$\nu \rightarrow $e+$\nu $ at
neutrino energies roughly in the range $0.3m_{e}\leq E_{\nu }\leq 3m_{e}$ 
\cite{Borexino_pp,Borexino_7Be,Borexino_pep}. In this region, the energy
dependence of the large mixing angle Mikheyev-Smirnov-Wolfenstein \cite%
{ms,lw,hs} (LMA-MSW) mixing is weak, and we take the NSI effects to be
concentrated in the source and detector, where we focus our attention in
this work.\footnote{%
Typically, works on solar NSI effects \cite{FLP-G,bolanos,g-g&m,klein}, have
included the whole solar spectrum and have taken the opposite point of view,
making detailed studies of the propagation effects and not including the
direct source and detector NSI contributions. For the completely opposite
view, where a solution of the solar neutrino problem was sought using \emph{%
only} NSI, in the spirit of \cite{lw}, see \cite{bergmann200}.} The
structure of the problem is quite similar to the problem of NSI effects in
very short-baseline reactor neutrino data, with the key difference that the
energy-independent, oscillation length-averaged solar propagation factor
brings in mixing parameters and carries source terms that are \emph{linear}
in the flavor-changing (FC) NSI parameters. This gives an advantage in
sensitivity over the very short-baseline case, where these NSIs appear only
quadratically.

In our formalism, described in Sec. 2, we combine source NSIs with detector
NSIs in a unified framework. To a first approximation, excellent in the $pp$
case, the oscillation probabilities in the low-energy region are
energy-independent, the result of averaging over the long propagation
distances. Since the source NSIs are included in the oscillation
probabilities, there is little energy dependence involving the NSI
parameters. On the other hand, the detector $\nu $-e cross sections depend
strongly on neutrino energy, \ including energy dependence due to the NSIs
at the detector. As a result, the convolution of the flux, the oscillation
probabilities and the cross sections over the neutrino energy spectrum
introduces energy dependence in the rate that is almost entirely due to the
detector. The resulting event rate treatment is outlined in Sec. 3. To
approximate the small $P_{ee}$ variation with energy expected from the
standard mixing model (SMM) and LMA-MSW model as $E_{\nu }$ rises to the $%
^{7}$Be and $pep$ sources, we evaluate the count rates with the electron
survival probability reduced by the appropriate factors relative to $pp$. As
our first application, we turn off the NSIs and fit $\sin ^{2}(\theta _{W})$
to the Borexino low-energy data in Sec. 4 and compare to the results of
several other studies \cite{amir_global, kmt2,vallewm}.

In our study of reactor short-baseline experiments \cite{amir_global, kmt2},
we showed that the data can not provide absolute bounds on the FC leptonic
NSI parameters involving $\bar{\nu}_{\mu }$ or $\bar{\nu}_{\tau }$, because
as the relevant \emph{source} NSI parameters approach zero, it leaves only
incident $\bar{\nu}_{e}$s, which cannot oscillate appreciably to $\bar{\nu}%
_{\mu }$ or $\bar{\nu}_{\tau }$ in the 10s of meters baselines like TEXONO 
\cite{deniz1}, LSND \cite{lsnd} or KARMEN \cite{karmen}. This is not true in
the present study, where the solar $\nu _{e}$s oscillate significantly to $%
\nu _{\mu }$s and $\nu _{\tau }$s, providing "wrong flavor" neutrinos at the
detector and subsequent bounds on the FC leptonic NSI parameters even when
the source FC NSIs are turned off. Because the oscillation and
flavor-changing NSI produce similar and sometimes mutually supporting
effects, the interesting problem of disentangling them arises. Those NSI
processes that do not involve oscillations such as those in very
short-baseline neutrino scattering and nuclear and particle decay
experiments then play an important role. We expand on this question upon
closing in Sec. VII.

The solar low-energy neutrino spectrum study is potentially ideal for the
study of NSI phases. As we noted in Refs. \cite{amir_global,kmt2}, reactor
data shows more sensitivity to these phases at the very low-energy end of
the neutrino spectrum. This is because the energy-dependent coefficients of
the phase-dependent terms, which are proportional to the electron mass m$%
_{e} $, becomes comparable to that of the other terms as the neutrino energy
becomes of the order of m$_{e}$ or less. This makes the \textit{pp}-chain
neutrino region below 1 MeV valuable for the NSI phase information. The
present work can be viewed as complementary to analyses of the TEXONO
reactor experiment \cite{deniz1,deniz2}, for example. Both experiments
involve a semi-leptonic process at the source and detection of a recoil
electron from an elastic neutrino-electron scattering at the detector. In
the solar case, $\nu _{e}$s are the beam neutrinos, while in the reactor
case $\bar{\nu}_{e}$s are the incident beam, and, as already noted, the
solar analysis requires oscillations while the reactor analysis does not. We
fit the source NSIs with solar data in Sec. 5, the detector NSIs in Sec. 6. 

As noted in Ref. \cite{Borexino_pp} if the Borexino precision is extended to
reach the 1\% level, it gives powerful leverage to determine the solar
metalicity and to explore neutrino properties. As dark-matter experiments
reach levels of sensitivity where solar background events become a problem,
in effect they become sensitive solar neutrino detectors \cite%
{solarDM,solarDM2,solarDM1,solarDM3} that can combine with future high
sensitivity multi-purpose neutrino experiments \cite%
{CJPL,HKexpt,JUNO,LENA,gmtvLena,SNO+} to make orders of magnitude increases
in the quality of low-energy solar physics data. We estimate the sensitivity
to NSI parameters that would be possible with these developments in Sec. 7,
commenting on the level of tightening of bounds on NSIs compared to those
following from the current data. In addition, we analyze source and detector
cross correlations among our future prospects possibilities and describe the
origins of several of the striking correlations that we find. We summarize
and conclude in Sec. 8. Our treatment of the $pp$ neutrino flux is given in
an Appendix.

\section{Formalism and Notation}

In this section we review all of the essential neutrino effective
Lagrangian, neutrino mixing and neutrino oscillation formulas in the
presence of non-universal (NU) and FC neutrino interactions, defining
relevant notation as we go.

\subsection{NSI effective Lagrangians at source and detector}

For the general setup under consideration, the sources of neutrinos and
antineutrinos are the nuclear fusion/decay processes in the sun, nuclear
beta decays in reactor cores or the pion decays at accelerators, while the
target particles at the detectors are electrons. Therefore the effective
four-Fermi Lagrangians governing the charged-current (CC) semi-leptonic
processes at the source \cite{kmt2,amir_global,kmt1,j&m1,TO_med_bl,j&m2,
b&g1} and the (anti)neutrinos-electron scattering processes \cite%
{kmt2,amir_global} at the detector are given as

\bigskip

\begin{eqnarray}
\mathcal{L}^{s} &=&\mathcal{L}_{NU}^{s}+\mathcal{L}_{FC}^{s} \\
\mathcal{L}^{\ell } &=&\mathcal{L}_{NU}^{\ell }+\mathcal{L}_{FC}^{\ell },
\end{eqnarray}%
where,%
\begin{eqnarray}
\mathcal{L}_{NU}^{s} &=&-2\sqrt{2}G_{F}\sum_{a,\alpha }(1+\varepsilon
_{\alpha \alpha }^{udL})(\bar{l}_{\alpha }\gamma _{\lambda }P_{L}U_{\alpha
a}\nu _{a})(\bar{d}\gamma ^{\lambda }P_{L}u)^{\dagger }+h.c., \\
\mathcal{L}_{FC}^{s} &=&-2\sqrt{2}G_{F}\sum_{a,\alpha \neq \beta
}\varepsilon _{\alpha \beta }^{udL}(\bar{l}_{\alpha }\gamma _{\lambda
}P_{L}U_{\beta a}\nu _{a})(\bar{d}\gamma ^{\lambda }P_{L}u)^{\dagger }+h.c.,
\\
\mathcal{L}_{NU}^{\ell } &=&-2\sqrt{2}G_{F}\sum_{\alpha }(\overline{e}\gamma
_{_{\lambda }}\left( \widetilde{g}_{\alpha R}P_{R}+(\widetilde{g}_{\alpha
L}+1)P_{L})e\right) (\bar{\nu}_{\alpha }\gamma ^{^{\lambda }}P_{L}\nu
_{\alpha }), \\
\mathcal{L}_{FC}^{\ell } &=&-2\sqrt{2}G_{F}\sum_{\alpha \neq \beta
}\varepsilon _{\alpha \beta }^{eP}(\bar{e}\gamma _{\lambda }Pe)(\bar{\nu}%
_{\alpha }\gamma ^{\lambda }P_{L}\nu _{\beta }).
\end{eqnarray}%
Here the superscript $s$ and $\ell $ designate semi-leptonic and purely
leptonic Lagrangians and the subscripts $NU$ and $FC$ correspond to the NU
and FC NSIs for both cases. $\alpha $\ and $\beta $ \ are the flavor-basis
indices and $a$ is mass-basis index\footnote{%
For simplicity, we consider only the left-handed quark helicity states and
do not include the right-handed terms in our discussion. The parameters $%
\varepsilon _{\alpha \beta }^{udL}$ are called $K_{\alpha \beta }$ in Ref. 
\cite{kmt1, kmt2}.}. The complex coefficients $\varepsilon _{\alpha \beta
}^{udL}$\ represent the relative coupling strengths of the flavor
combinations in the presence of new physics at solar, accelerator or reactor
sources and the complex coefficients $\varepsilon _{\alpha \beta }^{eP}$\
represent the relative coupling strengths of the flavor combinations in the
presence of new physics at the detector, while in the SM $\varepsilon
_{\alpha \beta }^{udL}$= 0 and $\varepsilon _{\alpha \beta }^{eP}$= 0. The
NU, flavor-diagonal interactions contain the NSI parameters $\varepsilon
_{\alpha \alpha }^{udL}$ at the source and both $\varepsilon _{\alpha \alpha
}^{eR}\ and$\ $\varepsilon _{\alpha \alpha }^{eL}$\ at the detector, where
these are implicitly given in the definitions of the coefficients $%
\widetilde{g}_{\alpha R}$\ and $\widetilde{g}_{\alpha L},$\ where$\ 
\widetilde{g}_{\alpha R}=\sin ^{2}\theta _{W}+\varepsilon _{\alpha \alpha
}^{eR}\ $and$\  \  \widetilde{g}_{\alpha L}=\sin ^{2}\theta _{W}-\frac{1}{2}%
+\varepsilon _{\alpha \alpha }^{eL}$. The hermiticity of the pure leptonic
effective Lagrangian, $L^{\ell },$\ requires that the detector NSI parameter
matrix is Hermitian and therefore, $\varepsilon _{\alpha \beta
}^{eR,L}=(\varepsilon _{\beta \alpha }^{eR,L})^{\ast }$, so the NU NSI
parameters are real, but the FC NSI parameters are complex in general. With
the effective Lagrangians defined, we turn next to the cross sections and
flux factors needed for the study of the NSI effects at the source and
detector.

\subsection{Neutrino oscillation probabilities at Earth}

For neutrinos at the low-energy end of the solar spectrum from $pp,^{7}Be$
and $pep$ reactions, the LMA-MSW expectation is that the mixing at Earth is
essentially the vacuum oscillation result. For example, Ref. \cite{altBe7}
makes this assumption and uses the Borexino $^{7}$Be data to bound leptonic
NSI parameters that enter the flavor-diagonal elastic $\nu $-e cross section
at the detector. We review the SMM case and the NSI contributions in this
section, returning to the small, low-energy LMA-MSW effects in the following
section.

\subsubsection{The standard mixing model result}

The oscillation amplitude takes the matrix form $A_{\alpha \beta }=U_{\alpha
a}X_{a}U_{a\beta }^{\dagger }$, where the flavor labels are Greek letters, $%
\alpha $ and $\beta $ in this case (summation over repeated indices is
implied). One can include NSIs in matrix form, such as $(1+\varepsilon
^{udL})U$ and the following argument still applies, because the averaging
involves only the mass basis indices $a,b,c...$. The $U$ matrix is the
neutrino mixing matrix for any number of neutrinos and the $X$ is the
diagonal phase matrix $X\ $= diag$(1,\exp (-i2\pi L/L_{21}^{osc}),\exp
(-i2\pi L/L_{31}^{osc},...)$. The oscillation length is defined as $%
L_{ab}^{osc}=4\pi E/(m_{a}^{2}-m_{b}^{2})$. The oscillation probability can
be written as

\begin{equation}
P_{\alpha \beta }=|A_{\alpha \beta }|^{2}=|U_{\alpha a}X_{a}U_{a\beta
}^{\ast }|^{2},
\end{equation}%
so the average over an oscillation length is then

\begin{equation}
\langle P\rangle _{\alpha \beta }=U_{\alpha a}U_{\beta a}^{\ast }U_{\alpha
a}^{\ast }U_{a\beta }=|U_{\alpha a}|^{2}|U_{\beta a}|^{2},  \label{eq:Pbar}
\end{equation}%
for the average over one cycle of the probability function. For example, the
electron survival averaged probability is $\langle P\rangle _{ee}$ = $%
(c_{12}c_{13})^{4}$ + $(s_{12}c_{13})^{4}$ + $s_{13}^{4}$, in the most
commonly used basis and notation \cite{PDG16}, where $s_{ij}\equiv \sin
\theta _{ij}$ and $c_{ij}\equiv \cos \theta _{ij}$.

\subsection{ The NSI effects at the source with oscillations}

For the case where there are NSIs only at the source, which in the solar
neutrino case$\ $means the semi-leptonic, or $\varepsilon ^{udL}$
parameters, we can write the matrix form of the amplitude A as%
\begin{equation}
A_{\alpha \beta }=[(1+\varepsilon ^{udL})UXU^{\dagger }]_{\alpha \beta }.
\end{equation}%
After averaging over an oscillation length to get the oscillation average
probability for our application, we have%
\begin{eqnarray}
\langle P\rangle _{\alpha \beta }^{NSI} &=&|[(1+\varepsilon
^{udL})U]_{\alpha a}|^{2}|U_{\beta a}|^{2}  \notag \\
&=&(|U_{\alpha a}|^{2}+2\func{Re}(U_{\alpha a}\varepsilon _{\alpha \gamma
}^{udL\ast }U^{\ast }{}_{\gamma a})+|\varepsilon _{\alpha \gamma
}^{udL}U_{\gamma a}|^{2})|U_{\beta a}|^{2}.
\end{eqnarray}%
The low-energy $\nu $ survival probability\footnote{%
When flavor-violating NSI are active, $\langle P\rangle _{ee}^{NSI}$ means
the probability that a neutrino produced with a positron at the source
arrives as a $\nu _{e}$ to produce a recoiled electron at the detector.}
from the solar core to Earth is then

\begin{eqnarray*}
\langle P\rangle _{ee}^{NSI} &=&(|U_{ea}|^{2}+2\func{Re}(U_{ea}\varepsilon
_{ee}^{udL\ast }U^{\ast }{}_{ea}+U_{ea}\varepsilon _{e\mu }^{udL\ast
}U^{\ast }{}_{\mu a}+U_{ea}\varepsilon _{e\tau }^{udL\ast }U^{\ast }{}_{\tau
a}) \\
&&+|\varepsilon _{ee}^{udL}U_{ea}|^{2}+|\varepsilon _{e\mu }^{udL}U_{\mu
a}|^{2}+|\varepsilon _{e\tau }^{udL}U_{\tau a}|^{2})|U_{ea}|^{2}.
\end{eqnarray*}%
Working to linear order in the FC $\varepsilon _{\alpha \beta }^{udL}$
parameters, the expression for the electron neutrino survival probability is
found to be 
\begin{eqnarray}
\langle P\rangle _{ee}^{NSI} &=&(1+2\func{Re}\varepsilon
_{ee}^{udL}+|\varepsilon _{ee}^{udL}|^{2})\langle P\rangle
_{ee}^{SMM}-(c_{23}\varepsilon _{-})c_{13}^{3}\sin 2\theta _{12}\cos 2\theta
_{12}  \notag \\
&&+(c_{23}\varepsilon _{+})(\frac{1}{2}c_{13}^{2}\sin 2\theta _{13}\sin
^{2}2\theta _{12}-\sin 2\theta _{13}\cos 2\theta _{13}),  \label{eq:linorder}
\end{eqnarray}%
where $\langle P\rangle _{ee}^{SMM}$ is the average standard oscillation
probability as given below Eq. (\ref{eq:Pbar}). The parameter combinations $%
c_{23}\varepsilon _{+}$ and $c_{23}\varepsilon _{-}$ are defined as 
\begin{eqnarray}
c_{23}\varepsilon _{+} &\equiv &\left \vert \varepsilon _{e\mu }^{udL}\right
\vert \cos (\phi _{e\mu }+\delta _{CP})s_{23}+\left \vert \varepsilon
_{e\tau }^{udL}\right \vert \cos (\phi _{e\tau }+\delta _{CP})c_{23}  \notag
\\
c_{23}\varepsilon _{-} &\equiv &\left \vert \varepsilon _{e\mu }^{udL}\right
\vert \cos \phi _{e\mu }c_{23}-\left \vert \varepsilon _{e\tau }^{udL}\right
\vert \cos \phi _{e\tau }s_{23},  \label{eq:defepspm}
\end{eqnarray}%
which are \textit{the} two observable FC parameters that appear at linear
order, which is why we focus on them in the present work as we did for the
medium-baseline experimental set up in Ref. \cite{kmt1}. As Eq. (\ref%
{eq:defepspm}) reminds us, when the NSI are present the CP violating phase $%
\delta $ of the SMM appears in $P_{ee}$ inextricably intwined with the the
NSI CP-violating phases \cite{j&m2,j&m3,g&CP,kmt1}. We will return to this
point in our discussion of the Borexino constraints on source NSI parameters
in Sec. V.

The coefficients of the NSI in all of the above equations involve the
oscillation mixing angles, which leads in some applications to ambiguities
between the roles of the two parameter sets \cite{kmt1, TO_med_bl}. This is
especially clear in the linear, FC terms in Eq. \ref{eq:linorder}. In our
numerical work we will set the mixing parameters to their central values in
the 2016 Particle Data Group review \cite{PDG16}. We return to this point in
Sec. VII, where we describe several ways that degeneracies can be
constrained and give an example.

\subsection{NSI effects at the solar neutrino detector}

The Lagrangian given in Eq. (2), leads to the following total cross sections
for the $\nu _{e}e-$ and $\nu _{\mu ,\tau }e-$scattering cases, similar to
those calculated in Ref. \cite{kmt2}, for $\bar{\nu}_{e}e-$ and $\bar{\nu}%
_{\mu ,\tau }e-$scattering, 
\begin{eqnarray}
\left[ \sigma (\nu _{e}e)\right] _{SM+NSI} &=&\frac{2G_{F}^{2}m_{e}}{\pi }%
T^{\max }[(\widetilde{g}_{eL}+1)^{2}+\underset{\alpha \neq e}{\Sigma }%
|\varepsilon _{\alpha e}^{eL}|^{2}  \notag \\
&&+\left( (\widetilde{g}_{eR})^{2}+\underset{\alpha \neq e}{\Sigma }%
|\varepsilon _{\alpha e}^{eR}|^{2}\right) \left( 1-\frac{T^{\max }}{E_{\nu }}%
+\frac{1}{3}\left( \frac{T^{\max }}{E_{\nu }}\right) ^{2}\right)  \notag \\
&&-\left( (\widetilde{g}_{eL}+1)\widetilde{g}_{eR}+\underset{\alpha \neq e}{%
\Sigma }\func{Re}[(\varepsilon _{\alpha e}^{eL})^{\ast }\varepsilon _{\alpha
e}^{eR}]\right) \frac{m_{e}T^{\max }}{2E_{\nu }^{2}}]  \label{eq:nue_xsec}
\end{eqnarray}%
and 
\begin{eqnarray}
\left[ \sigma (\nu _{\mu ,\tau }e)\right] _{SM+NSI} &=&\frac{2G_{F}^{2}m_{e}%
}{\pi }T^{\max }[\widetilde{g}_{\mu ,\tau L}^{2}+\underset{\alpha \neq \mu
,\tau }{\Sigma }|\varepsilon _{\alpha \mu ,\tau }^{eL}|^{2}  \notag \\
&&+\left( \widetilde{g}_{\mu ,\tau R}^{2}+\underset{\alpha \neq \mu ,\tau }{%
\Sigma }|\varepsilon _{\alpha \mu ,\tau }^{eR}|^{2}\right) \left( 1-\frac{%
T^{\max }}{E_{\nu }}+\frac{1}{3}\left( \frac{T^{\max }}{E_{\nu }}\right)
^{2}\right)  \notag \\
&&-\left( \widetilde{g}_{\mu ,\tau L~~}\widetilde{g}_{\mu ,\tau R}+\underset{%
\alpha \neq \mu ,\tau }{\Sigma }\func{Re}[(\varepsilon _{\alpha \mu ,\tau
}^{eL})^{\ast }\varepsilon _{\alpha \mu ,\tau }^{eR}]\right) \frac{%
m_{e}T^{\max }}{2E_{\nu }^{2}}],  \label{eq:numu_xsec}
\end{eqnarray}%
where 
\begin{equation}
\widetilde{g}_{\alpha R}=\sin ^{2}\theta _{w}+\varepsilon _{\alpha \alpha
}^{eR}\  \emph{and}\  \  \widetilde{g}_{\alpha L}=\sin ^{2}\theta _{w}-\frac{1}{%
2}+\varepsilon _{\alpha \alpha }^{eL},  \label{eq:defg}
\end{equation}%
and $m_{e}$ is the electron mass, $E_{\nu }$ is the neutrino energy and $%
T^{\max }$ is the maximum of the recoiled-electron energy in the detector, $%
\ T^{\max }(E_{\nu })\equiv E_{\nu }/(1+m_{e}/2E_{\nu })$, where $0<E_{\nu
}<0.420$ MeV for $pp$ events and $E_{\nu }$= 0.862 MeV and 1.44 MeV for $%
^{7} $Be and $pep$ events respectively.

The factors$\  \func{Re}[(\varepsilon _{\alpha e}^{eL})^{\ast }\varepsilon
_{\alpha e}^{eR}]$ and $\func{Re}[(\varepsilon _{\alpha \mu ,\tau
}^{eL})^{\ast }\varepsilon _{\alpha \mu ,\tau }^{eR}]$ can be written
equivalently as $|\varepsilon _{\alpha e}^{eL}||\varepsilon _{\alpha
e}^{eR}|\cos (\phi _{\alpha e}^{eL}-\phi _{\alpha e}^{eR})$ and $%
|\varepsilon _{\alpha \mu ,\tau }^{eL}||\varepsilon _{\alpha \mu ,\tau
}^{eR}|\cos (\phi _{\alpha \mu ,\tau }^{eL}-\phi _{\alpha \mu ,\tau }^{eR}),$
respectively, in terms of the magnitudes and CP-violating phases, $\phi
_{\alpha \beta }^{eL,R}$, \cite{j&m1,j&m2,j&m3,g&CP} of the complex, FC
parameters.

\bigskip

\section{\textit{pp}, $^{7}$\textit{Be} and \textit{pep} expected event
rates at Borexino}

The basic input for the event rate calculation requires the flux of
neutrinos at Earth, the number of electron targets in the fiducial volume of
the detector, the cross section for $\nu $+ e $\rightarrow \nu $ +e elastic
scattering\footnote{%
Unless indicated otherwise, we use the current Z-pole $\overline{MS}$ value
of $\sin ^{2}(\theta _{W})$ in cross section evaluations.} and the exposure
time. In Borexino the results are typically presented as a rate per 100 tons
of detector per day. The $pp$ result requires the convolution of the
relevant cross section and the $pp$ energy spectrum \cite{Borexino_pp,bks},
whereas the other two cases are line spectra and involve only the product of
total flux and the cross section at production energy. All the Borexino
low-energy papers \cite{Borexino_pp, Borexino_7Be, Borexino_pep} use the
high-metallicity total solar fluxes emitted as calculated by Serenelli,
Haxton and Pe$\widetilde{n}$a-Garay \cite{shp-g} for reference values in
their quoted "expected count rate" estimate in each case. We follow suit in
our rate estimates for $\chi ^{2}$ tests of parameter best-fits and limits. 
\footnote{%
The \textit{pep} rate uncertainties are not Gaussian \cite{Borexino_pep}.
Nonethless we use the nominal values given in Table I of Ref. \cite%
{Borexino_pep} to estimate the weight of the \textit{pep} data in our
various fits.}

\subsection{Estimate of matter effects on the $\protect \nu_{e}$ survival
probability}

The \textit{pp} continuous energy flux distribution of emitted neutrinos, $%
\phi (E_{\nu })_{pp}$, is outlined in the Appendix. For these lowest-energy
neutrinos ($E_{\nu }\leq $ 0.420 MeV), the matter effects on the probability 
$P_{ee}$ that a $\nu _{e}$ survives the trip from the Sun's core to the
detector are very small, less than a percent different from the
path-averaged, pure vacuum-mixing prediction. For the somewhat higher,
single energy $^{7}$\textit{Be }(0.862 MeV) and \textit{pep} (1.44 MeV)
neutrinos, the matter effects are still small, 4-5\%, but not entirely
negligible. For all three sources, we include the small corrections due to
matter effects\textbf{\ }to the pure vacuum value of $\langle P_{ee}\rangle $%
. The NSI matter effects on the \textit{pp} spectrum are negligible \cite%
{lopesNSI} and therefore not included. Because the NSI propagation
corrections to the LMA-MSW model \cite{VDK, BPW, Parke, PDG16} for P$_{ee}$\
will be small corrections to small effects below 2 MeV \cite{FLP-G} and our
CC semi-leptonic source NSI do not enter the matter effects, and, as with
the $pp$\ spectral shape, we do not include the remaining matter
contributions from the $\nu $\ -e forward scattering process in this paper%
\footnote{%
Using global data on neutrino propagation through Earth, neglecting $\nu $-e
NSI in matter and at detectors, Ref. \cite{g-g&m} finds that NC semileptonic
NSI improve the fit to the energy dependence of P$_{ee}$ in the 2 MeV to 10
MeV transition region.}.

The cross sections are those for neutrino-electron scattering defined in
Sec. II D, the number of electrons per 100 tons of target, $N_{e}$ = 3.307$%
\times 10^{31}$ and the $pp$ flux as summarized in the Appendix in Eq. (\ref%
{eq:flux_def}) and Eq. (\ref{eq:fluxfit}). For the $^{7}$Be and $pep$
fluxes, which have discrete energy spectra, we treat the fluxes as delta
functions in evaluating the rate in Eq. (\ref{eq:totrate}). Following
Borexino, we take the high-metallicity SSM flux values $\phi
_{7Be}=4.48\times 10^{9}$ cm$^{-2}$s$^{-1}$ at 0.862 MeV and $\phi
_{pep}=1.44\times 10^{8}$cm$^{-2}$s$^{-1}$ at 1.44 MeV to compute our
"expected values". To incorporate the mild energy dependence in the $^{7}$Be
and $pep$ cases, we use the analytic LMA-MSW matter dependence outlined in
the "Neutrino mixing" review in Ref. \cite{PDG16} for $P_{ee}$ and for $\cos
\theta _{12}$. As for the $pp$ case, we use electron density at average 
\textit{pp} neutrino production point in these expressions, determined by
taking the average production distance from the solar core \cite{lopes} and
then assuming an exponential decrease in density outward from the core in
the analytic approximations.

In summary, we find that the modifications to the straight energy-
independent vacuum value of $\langle P^{vac}\rangle _{ee}$ = 0.558 give the
values $\langle P^{pp}\rangle _{ee}$ = 0.554, $\langle P^{7Be}\rangle _{ee}$%
= 0.536 and $\langle P^{pep}\rangle _{ee}$= 0.529. We use these values in
our evaluation of production rates to compare to the Borexino values and to
set limits on parameters. The basic structure of the expected rate
calculations reads

\begin{equation}
R_{\nu }^{i}=N_{e}\int_{0}^{E_{max}}dE_{\nu }\phi ^{i}(E_{\nu })\left(
\sigma _{e}(E_{\nu })\langle P^{i}\rangle _{ee}+\sigma _{\mu ,\tau }(E_{\nu
})[1-\langle P^{i}\rangle _{ee}]\right) ,  \label{eq:totrate}
\end{equation}%
where $\langle P^{i}\rangle _{ee}$ are given in Eq. (\ref{eq:linorder}),
with the index \textit{i} indicating whether vac, \textit{pp}, $^{7}$Be or 
\textit{pep} is inserted for the factor $\langle P\rangle _{ee}^{SMM}$ in
the application. The cross sections $\sigma _{e}(E_{\nu })$ and $\sigma
_{\mu ,\tau }(E_{\nu })$ are defined in Eq. (\ref{eq:nue_xsec}) and Eq. (\ref%
{eq:numu_xsec}). The effects of the MSW-LMA model are small, but we find
that they do make noticeable difference in details of the fits. To test a
model where matter effects play no role at the lowest energies, one simply
adopts the energy-independent "vac" value for $\langle P\rangle _{ee}$.

\section{The Standard Model: low-energy fit to $\sin ^{2}(\protect \theta %
_{W})$}

There is ongoing interest in the low-energy determination of the weak-mixing
angle $\sin ^{2}(\theta _{W})$ \cite{vallewm, huberwm}. Presently the lowest
energy determination of the weak-mixing parameter sin$^{2}(\theta _{W})$ is
that provided by the parity-violation measurement in $^{133}$Cs at 2.4 MeV 
\cite{Cs,boulder,porsev}. Current and future solar neutrino measurements at
the $pep$ energy and below can test the sin$^{2}(\theta _{W})$ = 0.23867$\pm 
$0.00016 prediction of $\overline{MS}$ running of this parameter to the
sub-MeV region \cite{erler}.

With all NSI parameters set to zero in our theoretical rate Eq. (\ref%
{eq:totrate}), we fit $\sin ^{2}(\theta _{W})$ and determine its uncertainty
with the straightforward $\chi ^{2}$ estimater

\begin{equation}
\chi ^{2}(\sin ^{2}(\theta _{W}))=\sum_{i}\frac{(R_{\nu }^{i}(\sin
^{2}(\theta _{W}))-R_{exp}^{i})^{2}}{(\sigma _{exp}^{i})^{2}},
\label{eq:chi_ssq}
\end{equation}%
where $i$ runs over the solar neutrino sources \textit{pp}, $^{7}$Be and 
\textit{pep} and where the expression for the phenomenological rate is given
in Eq. (\ref{eq:totrate}). With all NSI parameters set to zero and the
PDG(2016) \cite{PDG16} value $\sin ^{2}(\theta _{W})$ = 0.2313, our expected
rate values are $R^{pp}$ = 132 (144$\pm $ 13 \cite{Borexino_pp}), $%
R^{^{7}Be} $ = 48.2 (46$\pm $1.5 \cite{Borexino_7Be}) and $R^{pep}$ = 2.85
(3.1$\pm $0.6 \cite{Borexino_pep}), where the Borexino measured values are
given in parentheses after each expected rate value. The expected value of
the rate is larger than the measured value in the $^{7}$Be case, so one
expects the fit will result in a smaller value of the weak-mixing angle.
Because the relative error is significantly smaller in this measurement, it
will have the largest impact on the combined fit.

Using Borexino's published values for the rates and their 1$\sigma $
statistical uncertainties for $pp$, $^{7}$Be and $pep$ direct detection, we
find a best-fit sin$^{2}(\theta _{W})$ = 0.224 $\pm $ 0.016, consistent with
both the $\overline{MS}$ value at the Z-boson mass and the low- energy
theoretical prediction \cite{erler}. As described above in Sec. III A, we
use the SMM with the LMA-MSW energy dependence as reviewed in Ref. \cite%
{PDG16} and include the effect of differing solar electron density at the
average production point for each $\nu _{e}$ source. Our result is also
consistent with values based on decay and reactor data studies \cite%
{porsev,kmt2,amir_global,deniz1, deniz2,vallewm}. Our central value and
uncertainty, which reflects only statistical fluctuations, are somewhat
smaller than those of the reactor data alone. \textit{With the inclusion of
the pp data, our value is below the energies of all other determinations of
the weak-mixing angle to date. } 

In Fig. 1, we show the $\Delta \chi ^{2}$ distribution for the three
individual spectra and for the combined fit. The 90\% C.L and 1$\sigma $
values are indicated by the dashed lines. The results of all the four cases
are given in Table I. Clearly, the fit is dominated by the $^{7}$Be data
with its 3\% uncertainty, compared to 9\% for \textit{pp} and 19\% for 
\textit{pep}, whose large uncertainty results in the obvious insensitivity
to $\sin ^{2}(\theta _{W})$ at very small values, where only the CC
contributes. In the \textit{pp} and $^{7}$Be cases, the insensitivity shows
up as a slight asymmetry in the limits, with the lower limits being
marginally weaker than the upper limits.

\begin{figure}[tbph]
\begin{center}
\includegraphics[width=5in]{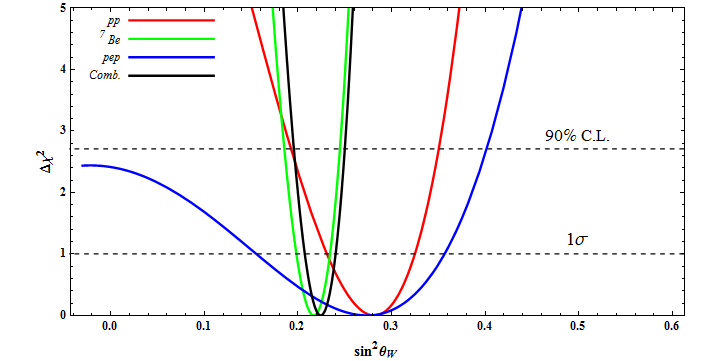}
\end{center}
\caption{\textbf{\ }The SM $\sin ^{2}(\protect \theta _{W}$) fit using the
solar low-energy spectrum of pp, $^{7}$Be, pep reaction measured by Borexino
experiment. From bottom to top, the 1$\protect \sigma $ and 90\%C.L. bands
are shown. The marked asymmetry in the bound from pep data is explained in
the text.}
\end{figure}

\begin{table*}[tbph]
\begin{center}
\begin{tabular}{l|l}
\hline \hline
Spectrum & Weak Mixing Angle \\ \hline
$pp$ & 0.281$\pm $0.047 \\ 
$^{7}$Be & 0.217$\pm $ 0.018 \\ 
pep & 0.274$\pm $0.101 \\ \hline
combined & 0.224$\pm $0.016 \\ \hline \hline
\end{tabular}%
\end{center}
\caption{ Shown are the $\sin ^{2}(\protect \theta _{W})$ fits to individual
and combined Borexino low- energy solar neutrino rates. Uncertainties shown
are averages of upper and lower values. The preference for a low value of $%
\sin ^{2}(\protect \theta _{W})$ results from the relatively small
uncertainty in the $^{7}$Be data. The text develops this point.}
\end{table*}

Another way to look at the the pattern of individual and joint fits is by
inspection of the $P_{ee}$ values determined by the Borexino data compared
to MSW-LMA expectation as summarized in "Extended Data Figure 2" in Ref. 
\cite{Borexino_pp}, which shows $P_{ee}$ vs. $E_{\nu }$. The $pp$ and $pep$
experimental points would like a larger value of $\sin ^{2}(\theta _{W})$,
increasing the cross section in the rate and permitting a smaller $P_{ee}$
while the $^{7}$Be point would like a smaller value, permitting a larger $%
P_{ee}$ value. The small error on the $^{7}$Be point gives it more weight,
and it pulls the fit down below the input value. All the data are within 1 $%
\sigma $ of the curve, so the effects are weak and our average final fit
value is consistent with the high energy precision value as well as with
other reported low-energy fits to decay and scattering data cited above.

\section{NSI at the source (Sun)}

The detector NSI are strongly correlated among themselves. The reasons lie
in the $P_{ee}^{NSI}$ probability dependence on the NSIs, Eq. (\ref%
{eq:linorder}), and in the structure of the event rate, Eq. (\ref{eq:totrate}%
), which shows the dependence of the rate on the electron survival
probability and the cross sections, Eqs. (\ref{eq:nue_xsec},\ref%
{eq:numu_xsec}), on the NSI parameters. When the $\varepsilon $ parameters
vary and change $\langle P_{ee}\rangle ,$ they change $1-\langle
P_{ee}\rangle $ in the opposite direction. For example, the coefficients of $%
\varepsilon _{+}$ and $\varepsilon _{-}$ in Eq. (\ref{eq:linorder}) are both
negative, with the latter's magnitude twice that of the former. When they
have the same sign, they lower the value of $P_{ee}^{NSI}$ as they grow and
raise it as they shrink. The value of $1-P_{ee}^{NSI}$ then compensates by
shrinking or growing. When the values of $\varepsilon _{+}$ and $\varepsilon
_{-}$ have opposite sign and grow in magnitude, they cancel each other and
tend to leave the rate unchanged, which leaves the $\chi ^{2}$ unchanged,
again leading to degeneracy. When combined with the variations of detector
NSI parameters, the situation improves, but long, narrow regions of
parameter space can still remain unbounded in some cases. The larger the
data set and/or the smaller the uncertainties in the data, the less the
impact of these degeneracies on the results of NSI searches.

NSI variations then make the system prone to the extended "filaments" of
same- likelihood regions. The large excursions to values greater than one
are a result of the rather large uncertainties in the data and our
truncation to linear order in the source parameters, which drops the
quadratic terms in the FC NSI and the overall normalization factor that
ensures that the total probability is constrained to 1.

The regions of parameter space in the neighborhood of the origin, captured
by the 1-parameter bounds, are useful as indications of the tightest
possible constraints with the given data. We restrict ourselves to these
regions for application of the Borexino \textit{pp}, $^7$Be and \textit{pep }
data to source NSI bounds in this section. In the presentation of future
prospects in Section VII we will return to the question of correlations
among NSI parameters.

Setting all of the NSI parameters at the detector equal to zero, we look at
the ranges of the $\func{Re}[\varepsilon _{ee}]$, $|\varepsilon _{ee}|$, $%
\varepsilon _{+}$ and $\varepsilon _{-}$ parameters allowed by
one-parameter-at-a-time fits to the experimentally measured values of the
Borexino event rates for \textit{pp}, $^{7}$Be and \textit{pep} solar
neutrino rates. The parameter $\func{Re}[\varepsilon _{ee}]$, the real part
of $\varepsilon $, enters linearly in the fit, which restricts its value
more tightly than its modulus, which enters quadratically. In effect the
modulus bound restricts the value of $\func{Im}[\varepsilon _{ee}]$, because
of the separate, tighter linear constraint on $\func{Re}[\varepsilon _{ee}]$%
. The constraints can then be presented separately in our results that
follow.

\begin{figure}[tbph]
\begin{center}
\includegraphics[width=5in]{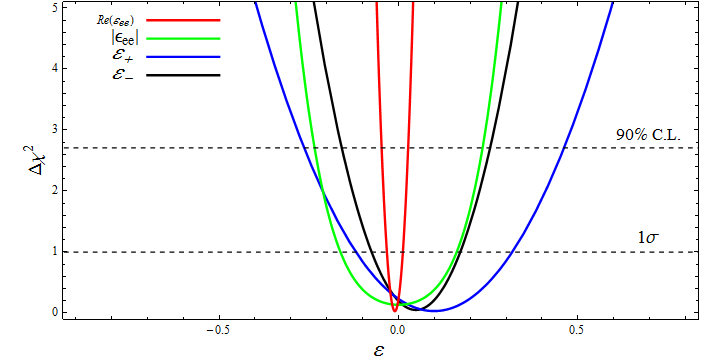}
\end{center}
\caption{Source NSI one parameter fits with the 1-$\protect \sigma $ or
68\%C.L. and 90\% C.L levels indicated by the horizontal dashed lines. The
corresponding best-fit values and 90\% C.L. spread is shown in Table II}
\label{fig:1&2paramfits}
\end{figure}

Figure \ref{fig:1&2paramfits} shows the results of all of the possible
one-parameter fits with the oscillation probability formula as given in Eq. (%
\ref{eq:linorder}) and Eq. (\ref{eq:defepspm}), where the independent NSI
parameters are $\func{Re}[\varepsilon _{ee}]$, $\left \vert \varepsilon
_{ee}\right \vert $, $\varepsilon _{+}$ and $\varepsilon _{-}$ 
in Fig. \ref{fig:1&2paramfits}. Figure \ref{fig:1&2paramfits} clearly shows
that the fit is about twice as sensitive to $\varepsilon _{-}$ as it is to $%
\varepsilon _{+}$, which reflects the fact that the coefficient multiplying $%
\varepsilon _{-}$ is about twice as large as that multiplying $\varepsilon
_{+}$ in Eq. (\ref{eq:linorder}).

With all other parameters set to zero, the 90\% C.L. one parameter bounds on
the source parameters are listed in Table \ref{tab:source_bounds}. These
generally compare reasonably well with bounds from other data and global
fits to 
\begin{table*}[tbph]
\begin{center}
\begin{tabular}{l|l|l|l|l}
\hline \hline
NSI Para. & $\func{Re}[\varepsilon _{ee}]$ & $\left \vert \varepsilon
_{ee}\right \vert $ & $\varepsilon _{+}$ & $\varepsilon _{-}$ \\ \hline
Best-Fits & $-0.0163$ & $0.0$ & $0.158$ & $0.082$ \\ 
Bounds & [-0.038, 0.038] & [-0.223,0.223] & [-0.370,0.370] & [-0.212,0.212]
\\ \hline \hline
\end{tabular}%
\end{center}
\caption{{}1-parameter source NSI parameter best fit values and their
corresponding 90\% C.L. uncertainties.}
\label{tab:source_bounds}
\end{table*}
solar data from other NSI studies. For example, comparing to Table II in
Ref. \cite{kmt2}, we find -0.72 $\leq |\varepsilon _{ee}|\leq $ 0.72 and
similarly for $\varepsilon _{e\mu }$ and $\varepsilon _{e\tau }$ at 90\%
C.L., while the corresponding allowed ranges given in Table \ref%
{tab:source_bounds} are about half that.

To make contact between bounds on $\varepsilon _{+/-}$ and bounds on the NSI
parameters $\varepsilon _{e\mu }$ and $\varepsilon _{e\tau }$, we display
the definitions from Eq. \ (\ref{eq:defepspm}) here.

\begin{eqnarray*}
c_{23}\varepsilon _{+} &\equiv &\left \vert \varepsilon _{e\mu }^{udL}\right
\vert \cos (\phi _{e\mu }+\delta _{CP})s_{23}+\left \vert \varepsilon
_{e\tau }^{udL}\right \vert \cos (\phi _{e\tau }+\delta _{CP})c_{23} \\
c_{23}\varepsilon _{-} &\equiv &\left \vert \varepsilon _{e\mu }^{udL}\right
\vert \cos \phi _{e\mu }c_{23}-\left \vert \varepsilon _{e\tau }^{udL}\right
\vert \cos \phi _{e\tau }s_{23}
\end{eqnarray*}%
The relationships are complicated by the dependence on the phases $\phi
_{e\mu }$, $\phi _{e\tau }$ and $\delta $, but simplify greatlly when $%
\delta $ = 0 or $\pi $. For example, when $\delta _{CP}$ = 0, only the real
parts of $\varepsilon _{e\mu /\tau }$ appear. Therefore the $\varepsilon
_{+} $ nonzero and $\varepsilon _{-}$ = 0 converts to the statement 
\begin{equation}
\func{Re}(\varepsilon _{e\mu })\approx \func{Re}(\varepsilon _{e\tau
})\approx \frac{1}{2}\varepsilon _{+},  \label{eq:del0}
\end{equation}%
and similarly when only $\varepsilon _{-}\neq 0$ and all other NSI = 0. The
approximate relationships become exact when $\tan (\theta _{23})$ = 1. In
short, the bounds on the $\varepsilon _{e\mu ,\tau }$ parameters are about
twice as tight as those listed for the $\varepsilon _{+/-}$ paramters . If
it turns out that $\delta \approx \frac{3\pi }{2}$, the current best-fit of
value of $\delta _{CP}$ in the SMM scenario, however, no such simple
connection can be made, since the $\varepsilon _{+}$ parameter is related to
the imaginary parts of the $\epsilon _{e\mu ,\tau }$ parameters. Nonetheless
we include the $\delta $ = 0 example to indicate that the constraints on one
set of parameters lead to similar constraints on the other.

The range allowed by early, model-independent surveys like \cite{biggio} are
significantly tighter than those we find from the current Borexino data,
however, where bounds set by CKM unitarity are $|\varepsilon _{ee}|\approx |%
\func{Im}[\varepsilon _{ee}]|\leq $ 0.041, and likewise for $\varepsilon
_{e\mu ,\tau }$. \footnote{%
Bounds on quark-neutrino NSI from matter effects alone, for example \cite%
{g-g&m,klein}, bound NC-type NSI and are not relevant to our CC NSI study.}
We note that the best-fit value for $\varepsilon _{-}$ implies a value 0.041
for $\varepsilon _{e\mu ,\tau }$, just at the 90\% C.L. limit quoted in \cite%
{biggio}. Our limits on this value are broad, but it is interesting to see
that a fit preferring this value is not ruled out. The bound quoted from 
\cite{biggio} is based on nuclear decay rates and is independent of the
values of the neutrino mixing parameters, whereas an input to our bound is
the value $\sin ^{2}\theta _{12}$=0.297 from Ref. \cite{PDG-2016}.

A detailed NSI study using Borexino precision $^{7}$Be data \cite{altBe7}
considers only the leptonic NSIs involved at the detector, assuming all
matter and source effects are negligible. We will comment on this study in
connection with our NSI at detector analysis next.

\section{NSI at the Borexino detector}

Here we turn to the sensitivity of the detector NSIs to the low-energy solar
data when all of the source NSIs are set equal to zero. Again, since the
matter effects in the LMA-MSW model are $4\%-5\% \ $at $E_{\nu }$ values
from the $^{7}$Be and $pep$ sources, we include this energy variation \cite%
{PDG16} in calculating the expected rates when NSIs at the detector are
active. As remarked earlier, we note that the NSI contributions to the
matter effects at low-energy are small \cite{FLP-G}, so the standard LMA-MSW
description suffices.\footnote{%
For a solar model independent check of the LMA-MSW, see Ref. \cite{VDK}.}

\subsection{NSIs in the $\protect \nu _{e}e-$scattering at the Borexino
detector}

\begin{figure}[tbph]
\begin{center}
\includegraphics[width=7in]{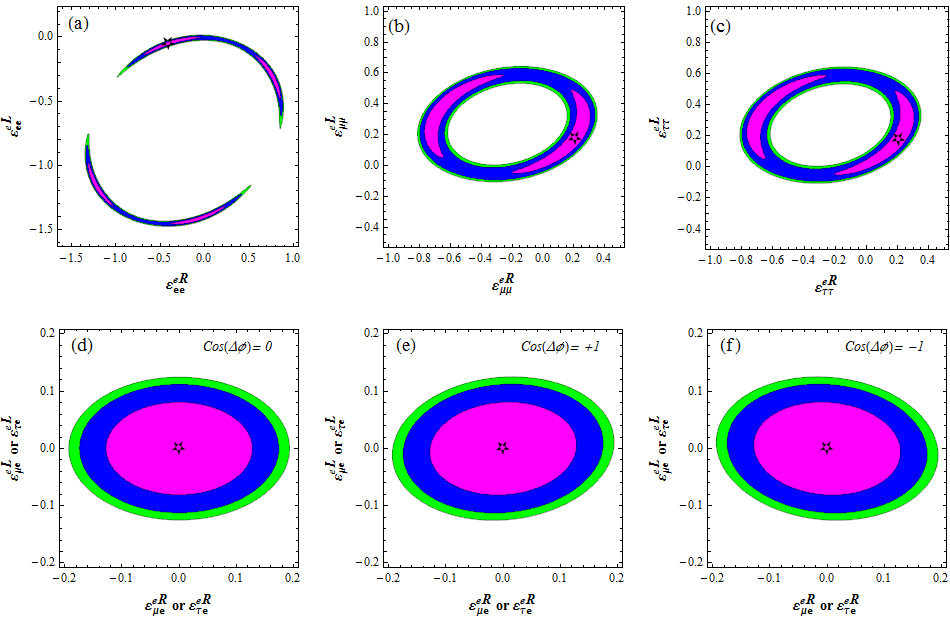}
\end{center}
\caption{Allowed 68\%(magenta), 90\% (blue) and 95\% (green) C.L. joint
parameter correlations of the NU and FC LH and RH subspaces for NSIs at the
detector. The star indicates the "best-fit" point for orientation in each
panel. The "no-NSI" point is at the origin, which is also the best fit point
for panels d, e and f. }
\label{fig:Alldetect}
\end{figure}

In this section we show and discuss the 68\%, 90\% and 95\% C.L. boundaries
of the independent combinations of detector NSI $\varepsilon _{\alpha \beta
}^{eL,R}$ parameters in Fig. \ref{fig:Alldetect}, where the stars show the
locations of the best-fit points closest to the no-NSI points at (0, 0) in
each case.\footnote{%
There are degenerate best-fit points in panels a, b and c in each of the
isolated, 1-$\sigma $ regions. The (0,0) no-NSIs' point is in or at the
boundary of the 1-$\sigma $ region in each case.} As expected, the data is
more sensitive to the left-handed (LH) NSI than to the right-handed (RH)
NSI, opposite to the case for reactor $\bar{\nu}_{e}$ fluxes, where the
roles of R and L are reversed compared to the solar $\nu _{e}$ flux \cite%
{kmt2,amir_global}. 

In the second row, the three cases correspond to the the NSI phase choices,
see Eq.\ref{eq:nue_xsec} and Eq.\ref{eq:numu_xsec}, $\cos (\phi _{\alpha
e}^{eL}-\phi _{\alpha e}^{eR})$ = 0, -1 and +1, from left to right. The
sense of the correlations in the cases $\cos (\phi _{\alpha e}^{eL}-\phi
_{\alpha e}^{eR})$ = +1/-1 are the same as in the $\bar{\nu}$ detection of
reactor neutrinos \cite{kmt2}, but the constraint on $\varepsilon _{\alpha
\beta }^{eL}$ is now tighter than that on $\varepsilon _{\alpha \beta }^{eR}$%
, since the roles of the dominant vs. subdominant terms in the cross section
are reversed. The relationship $\varepsilon _{\alpha \beta
}^{eR,L}=(\varepsilon _{\beta \alpha }^{eR,L})^{\ast }$, mentioned in Sec.
II A, and the dependence of the cross sections, Eq. \ref{eq:nue_xsec} and
Eq. \ref{eq:numu_xsec}, on only the moduli of the FC NSI parameters and on
the cosine of their phase differences lead to identical bounds for the $%
\varepsilon _{\mu ,\tau e}^{eL,R}$ and the $\varepsilon _{e\mu ,\tau
}^{eL,R} $ NSI parameters. We choose to display the bounds for the former
case in Fig. \ref{fig:Alldetect}. Likewise, the $\mu $ and $\tau $ labels
always appear symmetrically, so their figures are the same.

The two-parameter 90\% C.L. boundaries shown in Fig. \ref{fig:Alldetect}
contain one and two parameter bounds on the corresponding parameters, as
summarized in Table \ref{tab:detect_bounds}. 
\begin{table*}[tbph]
\begin{center}
\begin{tabular}{l|l|l|l|l}
\hline \hline
Sub-figure no. & \ 1-parameter (LH) & 1-parameter (RH) & 2-parameters (LH) & 
2-parameters (RH) \\ \hline
3(a) & $\varepsilon _{ee}^{eL}\in $[-0.017, 0.027] & $\varepsilon
_{ee}^{eR}\in $[-0.33, 0.25] & $\varepsilon _{ee}^{eL}\in $[-0.55,\ 0.02]$\ $
& $\varepsilon _{ee}^{eR}\in $[-0.80,\ 0.90] \\ 
3(b) & $\varepsilon _{\mu \mu }^{eL}\in $[-0.040, 0.04] & $\varepsilon _{\mu
\mu }^{eR}\in $[-0.10, 0.12] & $\varepsilon _{\mu \mu }^{eL}\in $[-0.61,
0.15] & $\varepsilon _{\mu \mu }^{eR}\in $[-0.33, 0.86] \\ 
3(d) & $\varepsilon _{\mu e}^{eL}\in $[-0.153, 0.153] & $\varepsilon _{\mu
e}^{eR}\in $[-0.238, 0.238] & $\varepsilon _{\mu e}^{eL}\in $[-0.152, 0.152]
& $\varepsilon _{\mu e}^{eR}\in $[-0.231, 0.231] \\ 
3(e) & $\varepsilon _{\mu e}^{eL}\in $[-0.152, 0.152] & $\varepsilon _{\mu
e}^{eR}\in $[-0.236, 0.236] & $\varepsilon _{\mu e}^{eL}\in $[-0.156,\ 0.156]
& $\varepsilon _{\mu e}^{eR}\in $[-0.244, 0.244] \\ 
3(f) & $\varepsilon _{\mu e}^{eL}\in $[-0.152, 0.152] & $\varepsilon _{\mu
e}^{eR}\in $[-0.236, 0.236] & $\varepsilon _{\mu e}^{eL}\in $[-0.156, 0.156]
& $\varepsilon _{\mu e}^{eR}\in $[-0.244, 0.244] \\ \hline \hline
\end{tabular}%
\end{center}
\caption{Parameter bounds from the detector-only study, with both
1-parameter and 2-parameter bounds given here at the 90\% C.L.. As noted in
the text, the 2-parameter bounds for cases (a), (b) and (c) lie partly
outside the range where they can be considered perturbations on the SMM,
they are given here for completeness. In all entries, the index $\protect \mu 
$ can be replaced by $\protect \tau $ and the \textit{e} and $\protect \mu $
subscripts interchanged.}
\label{tab:detect_bounds}
\end{table*}
Our leptonic NSI bounds listed in Table \ref{tab:detect_bounds} are
comparable to the comprehensive 90\% bounds for one parameter at-a-time from
Ref. \cite{amir_global}, Table 5, determined from global low-energy reactor
data . Compared to the bounds cited in \cite{mir&nun}, Table 3, the $%
\varepsilon _{ee}^{eL}$ and $\varepsilon _{\tau e}^{eL}$ bounds are tighter,
the $\varepsilon _{\tau e}^{eR}$ bound is the same and the $\varepsilon
_{ee}^{eR}$ and $\varepsilon _{\mu e}^{eL,R}$ bounds are weaker. Overall the
bounds we present are similar to the corresponding ones cited in Ref. \cite%
{mir&nun}.

For present discussion, there are no comparisons with the results of Ref. 
\cite{g-g&m}, which does not include the leptonic NSIs in its analysis, but
we can check against Ref. \cite{altBe7}. Their bounds are: $\varepsilon
_{ee}^{eL}$ [-0.046, 0.053], $\varepsilon _{ee}^{eR}$[-0.206, 0.157], $%
\varepsilon _{\tau \tau }^{eL}$[-0.231, 0.866], and $\varepsilon _{\tau \tau
}^{eR}$[-0.976, 0.726], which are weaker, stronger, weaker and weaker than
our corresponding bounds. These comparisons and others we present are meant
to be rough indicators of where our work stands in relation to other NSI
work, not a serious indication of "best bounds". The variation in the level
of treatment of theoretical and experimental uncertainties varies too much
among various quoted bounds to make direct, clear-cut comparisons among
published bounds.

We will see in the following section on future prospects that the degeneracy
effects are under control in most cases, though excursions into
unrealistically large perturbation regions, even at one sigma, are still
possible.

\section{Future prospects for sub-MeV determination of $\sin ^{2}(\protect%
\theta _{W})$ and for improved bounds on NSI parameters from low-energy
solar neutrinos}

A number of "Borexino-inspired" ideas for experiments have been advanced and
proposals made to measure \textit{pp} and other low-energy neutrinos to a
precision of 1\% or better. The objectives include determination of the
correct solar metallicity model and the corresponding fluxes of photons and
neutrinos, stringent testing of the LMA-MSW model of neutrino propagation in
the sun, and refined searches for exotic neutrino properties. A number of
dark-matter search proposals are reviewed in Ref. \cite{solarDM}, where the
emphasis is on their high efficiency for identifying and rejecting solar
neutrinos, thus providing a 1\%-3\% precision sample of solar neutrino data,
depending on the particular proposal. There are also dedicated solar
neutrino proposals such as Ref. \cite{CJPL} that aim for the same level of
precision. With these prospects in mind, we present estimates of the
improvement in sub-MeV measurement of $\sin ^{2}(\theta _{W})$ and NSI
parameter space boundaries that would follow from the improved precision.

\subsection{Determination of the low-energy solar value of $\sin ^{2}(%
\protect \theta _{W})$}

\begin{figure}[tbph]
\begin{center}
\includegraphics[width=5in]{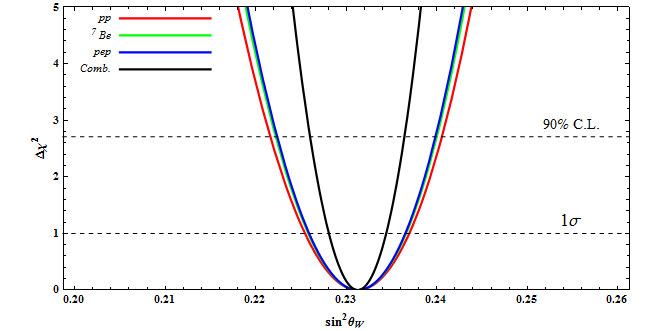}
\end{center}
\caption{\textbf{Future Prospects:} Low energy solar neutrinos fit to SM $%
\sin ^{2}(\protect \theta _{W}$). From bottom to top, the 1$\protect \sigma $
and 90\%C.L. bands are shown. The experimental rates are assumed to be those
of the standard model with $\sin ^{2}(\protect \theta _{W})$ set to the $%
\overline{MS}$ value at the Z-boson mass. At 5 $\protect \sigma $ the allowed
value of $\sin ^{2}(\protect \theta _{W})$ ranges up to 0. 2382. }
\label{fig:ssqth_fp}
\end{figure}

For the purpose of framing a test, we assume that the true value of the
weak-mixing angle is the high energy determination quoted in the PDG-2016
edition, 0.2313. With 1\% uncertainties in the measurements of \textit{pp}, $%
^{7}$Be and \textit{pep} rates, we ask at what level of confidence is it
consistent with the value 0.2387 predicted by SM renormalization group
running down to 10 -100 MeV \cite{erler,porsev} from the high energy
measured values? In Fig. \ref{fig:ssqth_fp}, we show the results of an
estimate of $\Delta \chi ^{2}$ values as a function of $\sin ^{2}(\theta
_{W})$. By eye one can see that the two values for $\sin ^{2}(\theta _{W})$
agree at a $\Delta \chi ^{2}$ of about 5, or about 98\% C.L.. Though our
estimate is rough, not including input parameter uncertainties, it indicates
that new, precision solar measurements have the potential to give useful
information about the value of this fundamental parameter at energies an
order of magnitude below those currently explored \cite{porsev}. 


\subsection{NSI at the source (Sun)}

\begin{figure}[tbph]
\begin{center}
\includegraphics[width=7in]{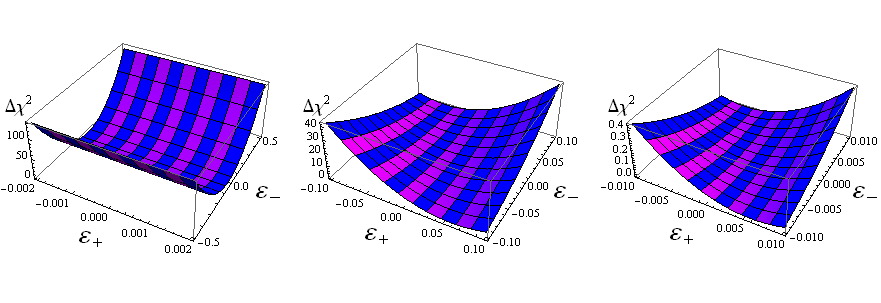}
\end{center}
\caption{\textbf{Future Prospects:} \textbf{3-dimensional view of the
correlation between} source NSI parameters $\protect \varepsilon _{+}$ and $%
\protect \varepsilon _{-}.$See text for details.}
\label{fig:3Dsource}
\end{figure}

Taking a 1\% nominal uncertainty on the experimental rate, and taking the
LMA-MSW model with neutrino mixing parameters at their PDG \cite{PDG16}
central values, we show the correlation between the $\varepsilon _{+}$ and $%
\varepsilon _{-}$ parameters over the limited ranges ($-0.10,+0.10$) in the
central panel of Fig. \ref{fig:3Dsource}. The left panel shows a range of $%
\varepsilon _{-}$ expanded by a factor of five and the range of $\varepsilon
_{+}$ narrowed to the vicinity around zero. This slice indicates the rapid
rise in $\Delta \chi ^{2}$ at small, fixed $\varepsilon _{+}$ and rising $%
\varepsilon _{-}$.

Focussing on the central figure, we see a degenerate trough of low $\Delta
\chi ^{2}$ values along a line starting from the right-front corner at about
($0.07,-0.1$) to the rear corner at about ($-0.03,0.1$), along which the
contributions from these two source NSIs tend to cancel each other, as
discussed in Sec. V. \ Similarly, the steep slopes along the line from the
front left corner at ($-0.1,-0.1$) to the rear right corner at ($0.1,0.1$)
show a region along which they add. The right panel shows this pattern blown
up in the center of the region covered by the central panel.

\begin{figure}[tbph]
\begin{center}
\includegraphics[width=5in]{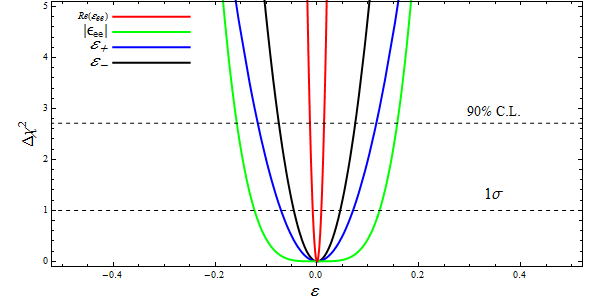}
\end{center}
\caption{\textbf{Future prospects:} Boundaries on individual source NSI
parameters $\protect \varepsilon _{+,-}$, $|\protect \varepsilon _{ee}|$, and $%
\func{Re}$\textbf{$(\protect \varepsilon _{ee})$} are shown with 1$\protect%
\sigma $ and 90\% C.L. lines. The latter are listed in Table IV}
\label{fig:FPsource}
\end{figure}

In Fig. \ref{fig:FPsource}, we show the individual bounds on all of the
source $\varepsilon $ parameters at 68\% and 90\%C.L.. This corresponds to
the $\varepsilon _{+}$= 0 slice of the left panel of Fig. \ref{fig:3Dsource}
in the range $-0.10$ $\leq \varepsilon _{-}\leq 0.10$. Fig. \ref%
{fig:FPsource} shows the same qualitative features as Fig. \ref%
{fig:1&2paramfits}, but the bounds on $\func{Re}$\textbf{$(\varepsilon
_{ee}) $}, $\varepsilon _{+}$ and $\varepsilon _{-}$ have been tightened by
factors of 4, as indicated in Table \ref{tab:FPsource}. As argued in Sec. V,
connecting these bounds with bounds on the FC NSI $\varepsilon _{e\mu }$ or $%
\varepsilon _{e\tau }$ suggests that the bounds shown in Table \ref%
{tab:FPsource} for $\varepsilon _{+}$ and $\varepsilon _{-}$ should be
divided by 2 for estimating the bounds on their flavor-labeled counterparts,
making them competitive with or tighter than those currently available in
Ref. \cite{biggio}, Table III, and Ref. \cite{mir&nun} ,Table IV, for
example.

\begin{table*}[tbph]
\begin{center}
\begin{tabular}{l|l|l|l|l}
\hline \hline
NSI Para. & $\func{Re}$\textbf{$(\varepsilon _{ee})\ $} & $|\varepsilon
_{ee}|$ & $\epsilon _{+}$ & $\epsilon _{-}$ \\ \hline
Best-Fits & $0.0$ & 0.0 & 0.0 & 0.0 \\ 
Bounds & $[-0.005,0.005]$ & $[-0.09,0.09$ & $[-0.05,0.05]$ & $[-0.023,0.023]$
\\ \hline \hline
\end{tabular}%
\end{center}
\caption{{}\textbf{Future Prospects:} 1-parameter at-a-time source NSI
parameter bounds at the 90\% C.L..}
\label{tab:FPsource}
\end{table*}

\subsection{NSI at the detector}

\begin{figure}[tbph]
\begin{center}
\includegraphics[width=7in]{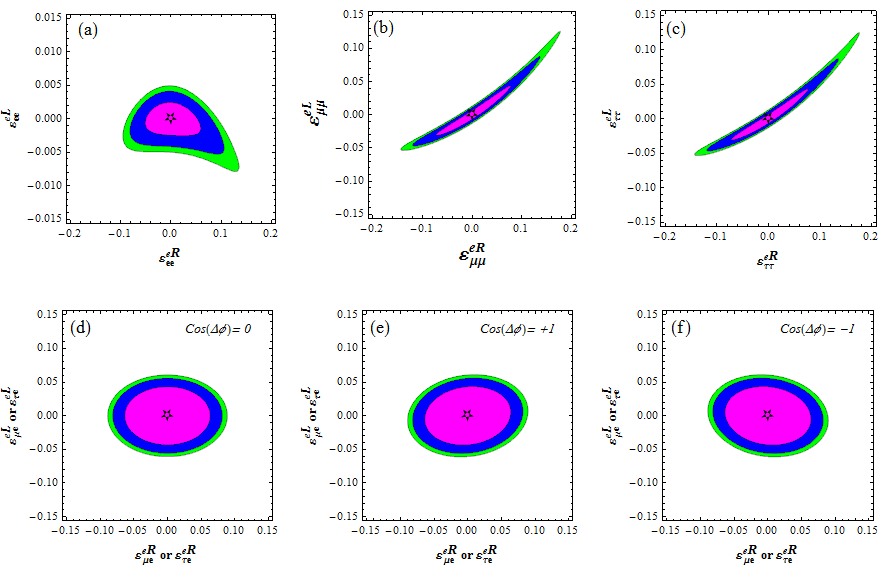}
\end{center}
\caption{\textbf{Future Prospects:} Allowed 68\% (magenta), 90\% (blue) and
95\% (green) C.L. NSI boundaries are shown for the detector-only case. By
definition of the "data" and the $\protect \chi ^{2}$, the best-fit value,
indicated by the star, is at the origin in each case.}
\label{fig:FPdetector}
\end{figure}

Following a similar procedure as described in Sec. VI, we explore the
two-parameter NSI subspaces and plot the corresponding two-dimensional space
boundaries in Fig. \ref{fig:FPdetector}. Again, the qualitative features are
similar to those of the contours shown in the Borexino data-based fits in
Fig. \ref{fig:Alldetect}, but the boundaries are tightened by factors of 4
to 5 in the NU case $\varepsilon _{ee}^{eL,R}$ and by factors of
two-to-three for the rest of the correlated pairs shown. The complete set of
bounds is summarized in Table \ref{tab:FPdetector}. 
\begin{table*}[tbph]
\begin{center}
\begin{tabular}{l|l|l|l|l}
\hline \hline
Sub-fig. no. & \ 1-parameter (RH) & 1-parameter (LH) & 2-parameters (RH) & 
2-parameters (LH) \\ \hline
7(a) & $\varepsilon _{ee}^{eR}\in $[-0.076, 0.084] & $\varepsilon
_{ee}^{eL}\in $[-0.0046, 0.0046] & $\varepsilon _{ee}^{eR}\in $[-0.085, 0.12]
& \multicolumn{1}{|l|}{$\varepsilon _{ee}^{eL}\in $[-0.0064, 0.0047]$\ $} \\ 
7(b) & $\varepsilon _{\mu \mu }^{eR}\in $[-0.02, 0.02] & $\varepsilon _{\mu
\mu }^{eL}\in $[-0.01, 0.01] & $\varepsilon _{\mu \mu }^{eR}\in $[-0.11,
0.12] & \multicolumn{1}{|l|}{$\varepsilon _{\mu \mu }^{eL}\in $[-0.042,
0.073]} \\ 
7(d) & $\varepsilon _{\mu e}^{eR}\in $[-0.112, 0.112] & $\varepsilon _{\mu
e}^{eL}\in $[-0.077, 0.077] & $\varepsilon _{\mu e}^{eR}\in $[-0.114, 0.114]
& \multicolumn{1}{|l|}{$\varepsilon _{\mu e}^{eL}\in $[-0.076, 0.076]} \\ 
7(e) & $\varepsilon _{\mu e}^{eR}\in $[-0.112, 0.112] & $\varepsilon _{\mu
e}^{eL}\in $[-0.077, 0.077] & $\varepsilon _{\mu e}^{eR}\in $[-0.117, 0.117]
& \multicolumn{1}{|l|}{$\varepsilon _{\mu e}^{eL}\in $[-0.080, 0.080]} \\ 
7(f) & $\varepsilon _{\mu e}^{eR}\in $[-0.112, 0.112] & $\varepsilon _{\mu
e}^{eL}\in $[-0.077, 0.077] & $\varepsilon _{\mu e}^{eR}\in $[-0.117, 0.117]
& \multicolumn{1}{|l|}{$\varepsilon _{\mu e}^{eL}\in $[-0.080, 0.080]} \\ 
\hline \hline
\end{tabular}%
\end{center}
\caption{\textbf{Future Prospects}: Parameter bounds are listed for the
detector-only study. $\protect \tau $ can be substituted for $\protect \mu $
in all entries to obtain the corresponding $\protect \tau $ limits. Both
1-parameter and 2-parameter bounds are given here at the 90\% C.L..}
\label{tab:FPdetector}
\end{table*}

Comparing the limits illustrated in Fig. \ref{fig:FPdetector} and detailed
in Table \ref{tab:FPdetector} with those in Eq. (45) of Ref. \cite{biggio}
or with the compilation presented in Table 3 of Ref. \cite{mir&nun}, we find
that in every case our future prospects estimates of detector NSI bounds are
up to an order of magnitude tighter than those listed in these reviews.
Comparing with the global analysis of short-baseline neutrino results in
Ref. \cite{amir_global}, we also find that, with the exception of the $%
\varepsilon _{ee}^{eR}$ case, our estimates of the possible extension of the
search for new physics go significantly deeper with 1\% solar data.

\subsection{Correlation of source and detector NSI parameters}

\begin{figure}[tbph]
\begin{center}
\includegraphics[width=7in]{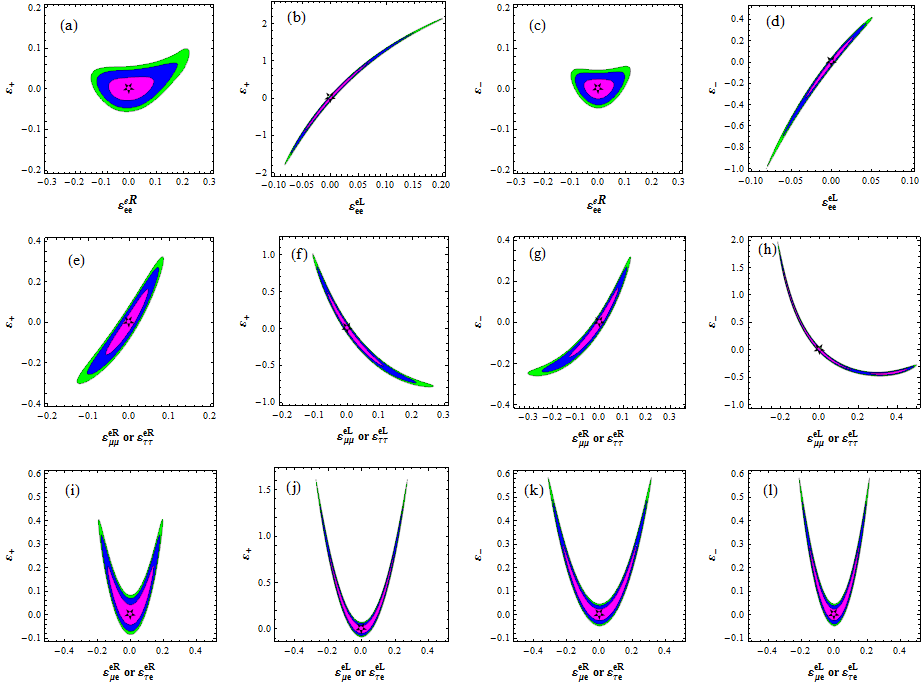}
\end{center}
\caption{\textbf{Future Prospects:} Future prospect boundaries on combined
source and detector NSI parameter correlations at 68\% (magenta), 90\%
(blue), and 95\% (green). For the detector FC NSI parameters from (i) to
(l), the C.L. regions of $\protect \varepsilon _{e\protect \mu }^{eR}$ etc.
are the same as those of the parameters shown because of the hermiticity of
the leptonic Lagrangian.}
\label{fig:FPsource&detector}
\end{figure}

To find correlations between the source and detector NSIs in the case of
proposed 1\% precision experiments, we study the two- parameter subspaces
displayed in Fig. \ref{fig:FPsource&detector}. These results show a striking
difference between the results for the case of the oscillating long baseline
solar experiments and the very short-baseline TEXONO-type experiments. The
source vs. detector plots and parameter bound tables contain strong
one-parameter limits and strong two-parameter limits on the detector FC NSI
in the region of small source parameters, shown in panels (i) through (l),
while the TEXONO type experiments leave this region unbounded, as indicated
in Fig. 2 of Ref. \cite{kmt2}. The reason is that only direct emission of $%
\bar{\nu}_{\mu }$ or $\bar{\nu}_{\tau }$ from the source can lead to signals
in the detector from $\bar{\nu}_{\mu }$-e interactions in the latter case,
so turning off the NSIs producing $\bar{\nu}_{\mu }$ at the source
eliminates the bound on FC events at the detector. In the solar oscillation
case, $\nu _{e}$ from the source can oscillate to $\nu _{\mu }$ and
contribute signal from these strictly NC type interactions.

Because the effective Lagrangian, Eq.(6), is Hermitian and the cross
sections in Eq. (\ref{eq:nue_xsec}) and Eq. (\ref{eq:numu_xsec}) contain
only absolute magnitudes of the FC NSIs and the cosine of the difference of
their phases, interchanging flavor labels $\mu $ and e leaves the results in
the figures and in the tables unchanged. Again the $\tau $ and $\mu $
results are all the same, so only the latter cases are shown.

The source and detector correlation plots show the features that, as with
the source-only study, there are stronger constraints on $\varepsilon _{-}$
than on $\varepsilon _{+}$ and on $\varepsilon _{e\alpha }^{eL}$ than on $%
\varepsilon _{e\alpha }^{eR}$. They have the distinctive feature of "wings"
that are long compared to the compact region around the origin, where the
NSI are confined to small deviations from NSIs = 0. These tenuous regions of
parameter space allow relatively long, highly correlated, excursions from
the LMA-MSW picture at 90\% C.L. in many cases. These arise from some
surprisingly tight correlations between the $\varepsilon _{\alpha \beta
}^{eR,L}$ contributions to the cross sections at the detector and the $%
\varepsilon _{+,-}$ contributions to $P_{ee}^{NSI}$ from the solar fusion
source. For example, take panels 8(b) and 8(f). In panel 8(b), as $%
\varepsilon _{ee}^{eL}$ grows it drives growth in $\sigma (\nu _{e}e)$,
while growth in $\varepsilon _{+}$ drives contraction of $P_{ee}$. The
product $P_{ee}\times \sigma (\nu _{e}e)$ appears in the rate equation, so
there is evidently a highly tuned line of correlated values that allows a
long, thin excursion from the best-fit value at the origin to large values
of this NSI pair, while still staying within the prescribed confidence
levels. A similar situation arises in panel 8(f), but here as $\varepsilon
_{\mu \mu }^{eL}$ grows it makes $\tilde{g}_{\mu L}$ shrink, which causes $%
\sigma (\nu _{\mu }e)$ to shrink. If $\varepsilon _{+}$ grows, $P_{ee}$
shrinks, but then 1-$P_{ee}$ grows and compensates for the shrinking of the $%
\nu _{\mu }$-e cross section, leading to an anti-correlation as shown in
panel 8(f). The situation is similar in the corresponding $\varepsilon _{-}$
plots 8(d) and 8(h).

In Table \ref{tab:FPs&d_cor}, we display the complete lists of source and
detector two-parameter correlated bounds. The single parameter bounds are
the same as found in the corresponding source-only and detector-only
studies, as they must be, and the future potential comparisons with
published bounds and related discussion presented there applies here, namely
that the single parameter bounds show potential for significant improvement
in sensitivity to new physics. The range of the correlated 90\% C.L. bounds
on all of the purely leptonic NSI, those that affect the cross sections that
apply at the Borexino detector, are rather tight, consistent with small
deviations from the SM values and generally show promise to tighten the
bounds listed in TABLE V of Ref. \cite{amir_global}. The correlated bounds
(a), (c), (e), (g), (i), (k) and (l) on $\varepsilon _{+,-}$ are consistent
with the linear approximation assumed in Sec. II C and applied throughout.
Even in these cases, however, the correlations with the leptonic NSIs carry
the limits well beyond their single parameter ranges.

In the cases (b), (d), (f), (h) and (j), the ranges of the $\varepsilon
_{+,-}$ parameters allowed at 90\% C.L. reach well beyond linear
approximation values. To treat them consistently requires that the $P_{ee}$
expressions complete to quadratic order be employed. This goes beyond the
scope of the present work, which takes a first look at the role of NSIs at
the source in the low-energy range of solar neutrinos, where matter effects
are very small, as our estimates show, and the semi-leptonic NSIs that apply
to the \textit{pp}, $^{7}$Be and \textit{pep} processes are competitive.

\begin{table*}[t]
\begin{center}
\begin{tabular}{l|l|l|l|l}
\hline \hline
Fig. No. & \ 1-parameter(RH) & 1-parameter(LH) & 2-parameters(RH) & 
2-parameters(LH) \\ \hline
\multicolumn{1}{c|}{8(a)} & $\varepsilon _{ee}^{eR}\in $[-0.07, 0.07] & $%
\varepsilon _{+}\in $[-0.07, 0.07] & $\varepsilon _{ee}^{eR}\in $[-0.08,
0.11] & $\varepsilon _{+}\in $[-0.07, 0.09] \\ 
\multicolumn{1}{c|}{8(b)} & $\varepsilon _{ee}^{eL}\in $[-0.004, 0.004] & $%
\varepsilon _{+}\in $[-0.07, 0.07]$\ $ & $\varepsilon _{ee}^{eL}\in $%
[-0.075, 0.15] & $\varepsilon _{+}\in $[-1.5, 1.8]$\ $ \\ 
\multicolumn{1}{c|}{8(c)} & $\varepsilon _{ee}^{eR}\in $[-0.08, 0.08] & $%
\varepsilon _{-}\in $[-0.04, 0.04] & $\varepsilon _{ee}^{eR}\in $[-0.09,
0.10] & $\varepsilon _{-}\in $[-0.04, 0.04] \\ 
\multicolumn{1}{c|}{8(d)} & $\varepsilon _{ee}^{eL}\in $[-0.003, 0.003] & $%
\varepsilon _{-}\in $[-0.03, 0.03] & $\varepsilon _{ee}^{eL}\in $[-0.06,
0.043] & $\varepsilon _{-}\in $[-0.7, 0.37] \\ 
\multicolumn{1}{c|}{\ 8(e)} & $\varepsilon _{\mu \mu }^{eR}\  \in $[-0.022,
0.022] & $\varepsilon _{+}\in $[-0.07, 0.07] & $\varepsilon _{\mu \mu
}^{eR}\in $[-0.1, 0.07] & $\varepsilon _{+}\in $[-0.26, 0.27] \\ 
\multicolumn{1}{c|}{8(f)} & $\varepsilon _{\mu \mu }^{eL}\  \in $[-0.010,
0.010] & $\varepsilon _{+}\in $[-0.07, 0.07] & $\varepsilon _{\mu \mu
}^{eL}\in $[-0.10, 0.21] & $\varepsilon _{+}\in $[-0.75, 0.85] \\ 
\multicolumn{1}{c|}{8(g)} & $\varepsilon _{\mu \mu }^{eR}\  \in $[-0.006,
0.006] & $\varepsilon _{-}\in $[-0.009, 0.009] & $\varepsilon _{\mu \mu
}^{eR}\in $[-0.27, 0.34] & $\varepsilon _{-}\in $ [-0.27, 0.34] \\ 
\multicolumn{1}{c|}{8(h)} & $\varepsilon _{\mu \mu }^{eL}\  \in $[-0.010,
0.010] & $\varepsilon _{-}\in $ [-0.04, 0.04] & $\varepsilon _{\mu \mu
}^{eL}\in $[-0.21, 0.49] & $\varepsilon _{-}\in $ [-0.47, 1.8] \\ 
\multicolumn{1}{c|}{$\ $8(i)} & $\varepsilon _{\mu e}^{eR}\  \in $[-0.08,
0.08] & $\varepsilon _{+}\in $[-0.07, 0.07] & $\varepsilon _{\mu e}^{eR}\in $%
[-0.18, 0.18] & $\varepsilon _{+}\in $ [-0.07, 0.34] \\ 
\multicolumn{1}{c|}{8(j)} & $\varepsilon _{\mu e}^{eL}\in $[-0.06, 0.06] & $%
\varepsilon _{+}\in $[-0.06, 0.06] & $\varepsilon _{\mu e}^{eL}\in $[-0.25,
0.25] & $\varepsilon _{+}\in $[-0.07, 1.3] \\ 
\multicolumn{1}{c|}{\ 8(k)} & $\varepsilon _{\mu e}^{eR}\  \in $[-0.08, 0.08]
& $\varepsilon _{-}\in $[-0.04, 0.04] & $\varepsilon _{\mu e}^{eR}\  \in $%
[-0.29, 0.29] & $\varepsilon _{-}\in $[-0.04, 0.49] \\ 
\multicolumn{1}{c|}{8(l)} & $\varepsilon _{\mu e}^{eL}\  \in $[-0.055, 0.055]
& $\varepsilon _{-}\in $[-0.040, 0.040] & $\varepsilon _{\mu e}^{eL}\  \in $%
[-0.20, 0.20] & $\varepsilon _{-}\in $[-0.04, 0.48] \\ \hline \hline
\end{tabular}%
\end{center}
\caption{\textbf{Future Prospects:} Parameter bounds from the source and
detector correlation study. In each entry where $\protect \mu $ appears, $%
\protect \tau $ can be substituted with the same range of values for the
bounds. Both 1-parameter and 2-parameter bounds are given here at the 90\%
C.L..}
\label{tab:FPs&d_cor}
\end{table*}

\subsection{Correlations among NSI and oscillation parameters}

The neutrino-electron cross sections themselves involve the oscillation
parameters only indirectly through the rate calculation, so the focus is on
the source parameters. These are directly entangled with the mixing
parameters through the $\nu _{e}$ arrival factor $\langle P\rangle
_{ee}^{NSI}$ in Sec. IIC. For our case, $\theta _{23}$ is involved only in
the definitions of $\varepsilon _{+}$ and $\varepsilon _{-}$, which relate
them to $\varepsilon _{e\mu }$ and $\varepsilon _{e\tau }$, so it ties only
indirectly to other experiments. The mass-squared difference $%
m_{2}^{2}-m_{1}^{2}$ enters only in the small, matter-effect corrections and
our results are insensitive to this parameter. Moreover, in reactor neutrino
applications such as the JUNO experiment, where semileptonic NSI are
involved at both source and detector, the fits to simulated data that
include NSI show little impact on $m_{2}^{2}-m_{1}^{2}$ best fits \cite%
{TO_med_bl}.

The situation is quite different for $\theta _{12}$ and $\theta _{13}$,which
largely determine the coefficients of $\varepsilon _{-}$ and $\varepsilon
_{+}$, leading to ambiguity in the interpretation of data in terms of
neutrino masses and mixing if NSI at source and/or detector are active. In
the case of reactor experiments, this effect \cite{grossman95} has been
studied in \cite{OZ,kmt1, kmt2,TO_med_bl}, as well as in long baseline
experiments \cite{blennow_lng_bl,ANK} where NSI matter effects could also be
important. Because there are relatively model independent bounds from
nuclear and particle decay and from very short-baseline experiments, there
are bounds on source and detector NSI of $\leq $ 0.05 on most of the
parameters \cite{biggio}. These are independent of oscillation phenomena,
which strongly limits the range of possible NSI induced error in determining
the values of the neutrino mass and mixing parameters.

There is generally expected to be ambiguity and apparent tension among
measurement of the basic neutrino mass and mixing parameters, possibly owing
to NSI parameters. Global fits that involve both parameter sets are
certainly warranted at each stage of advance in experimental scope and and
precision. Our aim here has been to bring out the potential for low-energy
solar neutrino physics to play a useful part in these future analyses.

\section{Summary and conclusions}

As emphasized in the Introduction, we have focused on the use of low-energy, 
$pp$, $^{7}$Be and $pep$ Borexino direct observation results, published
between 2011 and 2014 \cite{Borexino_pp, Borexino_7Be, Borexino_pep}, to
explore possible effects due to NU and FC NSIs. The primary motivation was
the relative insensitivity of the survival probability P$_{ee}$ to matter
effects, in the LMA-MSW picture, with or without effects due to NSI in
neutrino-matter forward elastic scattering. This feature, within reasonable
approximations, permits the focus to be on the direct NSI effects entering
at the fusion process in the sun and the modifications to $\nu $-e
interactions at the detector.

After introducing our formulation and basic notation in Sec. II, we outlined
the model calculation of neutrino interaction rates in Borexino in Sec. III.
Then in Sec. IV we set all NSI parameter values to zero to find the best-fit
value and allowed 1$\sigma $ and 90\% C.L. ranges of $\sin ^{2}(\theta _{W})$%
, extending the determination of the weak angle to the sub MeV energy
region. We found $\sin ^{2}(\theta _{W})$ = 0.224 $\pm $0.016, consistent
with the $\overline{MS}$ PDG value 0.23126(5) at the Z-boson mass,
renormalization group running values from the 100 GeV to 10 MeV range,
0.23867$\pm $ 0.00016 \cite{erler}, 0.2381(6) \cite{porsev}, and with
another recent low-energy study \cite{vallewm}. Our estimate in Sec. 7 of
the improvement in uncertainty possible with 1\% solar data shows that the
predicted value running to low-energy from the value at the Z-pole can be
tested at about 98\% C.L., but may not be sufficient to make a decisive
determination.

In Secs. V, VI and VII we then reported and discussed the results of our $%
\chi ^{2}$ analysis of the measured rates vs. the modeled rates, looking
systematically at allowed parameter boundaries of single and joint parameter
choices for source alone, detector alone and combined source and detector.
The results showed consistency\footnote{%
The consistency is in the sense that some bounds are stronger, some are
weaker and some are essentially the same.} with other solar studies that
focussed on matter effects \cite{g-g&m}, or on NU NSI at the detector alone 
\cite{altBe7}, or on short-baseline reactor data with both NU and FC NSI
effects at source and detector \cite{kmt2,amir_global}.

In Sec. VII we studied the impact that an improvement of the uncertainty in
experimental rates to the 1\% level, as targeted in a number of proposals
for new generation solar neutrino or dark-matter experiments, where the
solar neutrino background is recognized as a serious problem for experiments
hoping to increase dark-matter search sensitivities to new levels. As
expected, we found order of magnitude increases in sensitivity in some one
parameter at-a-time bounds, though the two-parameter space source vs.
detector level of improvements are mixed, due to largely to the high degree
of correlation among some of the parameters. This is evidence of systematic
compensation among terms in the fit functions, which leads to limited growth
of overall sensitivity. Developing tests that break these degeneracies goes
beyond the goal of this work to take a first look at bounds on source NSI
using the recent Borexino low-energy data and to revisit bounds on detector
NSI and to make a survey of the correlations between the two.

We conclude that current \textit{pp}, $^{7}$Be and \textit{pep} neutrino
rate measurements help narrow the range of lepton flavor violating NSI in
both semi-leptonic and leptonic NSI cases and that future 1\% measurements
will greatly improve the search for new physics effects. At the same time,
we have identified some strong correlations among NSI parameters and
ambiguities between oscillation and NSI effects that make bounding some of
the parameters a challenge and an important goal for future work.

The low-energy end of the solar neutrino spectrum will continue to be of
great interest for decades to come for reasons of straight neutrino physics,
solar physics and dark-matter physics. Our study showed the complementarity
of NSI source and detector low-energy solar data analysis to other NSI
searches. We found indications of limitations in our straightforward $\chi
^{2}$ treatment in the case of the source - detector cross correlation
studies, which brings out the need to explore more comprehensive analyses to
take full advantage of future experimental precision data.

\section{Appendix}

\subsection{\textbf{The $pp$ neutrino flux}}

\begin{figure}[tbph]
\begin{center}
\includegraphics[width=5in]{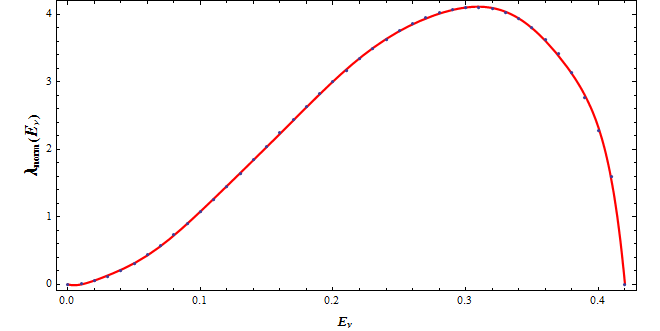}
\end{center}
\caption{Normalized $pp$\ spectrum $d\protect \lambda /dE_{\protect \nu }$
from Table IV of Ref. \protect \cite{b&u1}}
\label{fig:Keps1}
\end{figure}

We show the normalized $pp$ electron neutrino spectrum based on Table IV
from Ref. \cite{b&u1} in Fig. \ref{fig:Keps1}. As we mentioned in Sec. III
A, the NSI forward scattering effects on the \textit{pp} spectrum are very
small \cite{lopesNSI}, and we do not include them here. 
The dots in Fig. \ref{fig:Keps1} are the SSM points of the \textit{pp}
normalized spectrum, and the red line is the curve of the fit function, Eq. (%
\ref{eq:fluxfit}). We assume the total flux to be $\Phi _{pp}$ = 5.98$\times
(1\pm 0.006)\times 10^{10}$ cm$^{-2}$ s$^{-1}$, corresponding to the
high-metallicity model used in Borexino's publication \cite{Borexino_pp} to
calculate their expected rate of 131 $\pm $ 2. The flux as a function of
energy is then 
\begin{equation}
\phi (E_{\nu )})_{pp}=\Phi _{pp}\times d\lambda (E_{\nu })/dE_{\nu },
\label{eq:flux_def}
\end{equation}%
where the normalized flux energy distribution is fit by the power series, 
\begin{equation}
\frac{d\lambda (E_{\nu })}{dE_{\nu }}|_{fit}=\sum_{1}^{11}a_{n}(E_{\nu
})^{n-1},  \label{eq:fluxfit}
\end{equation}%
with the unit normalization maintained to one part in $10^{4}$. The fitting
parameters are given in Table \ref{tab:fluxfit}.

\begin{table*}[tbph]
\begin{center}
\begin{tabular}{|l|l|l|l|l|l|}
\hline
$a_{1}$ & $-6.21914$ & $a_{5}$ & $-6.82779.10^{6}$ & $a_{9}$ & $%
-1.39822.10^{9}$ \\ \hline
$a_{2}$ & $835.245$ & $a_{6}$ & $5.06675.10^{7}$ & $a_{10}$ & $%
1.49676.10^{9} $ \\ \hline
$a_{3}$ & $-28352.1$ & $a_{7}$ & $-2.41275.10^{8}$ & $a_{11}$ & $%
-6.91255.10^{11}$ \\ \hline
$a_{4}$ & $573193$ & $a_{8}$ & $7.3743.10^{8}$ & $-$ & $-$ \\ \hline
\end{tabular}%
\end{center}
\caption{The coefficients $a_{i}$ for the fit to the data from Table IV of
Ref. \protect \cite{b&u1}}
\label{tab:fluxfit}
\end{table*}

\begin{acknowledgments}
DWM thanks the Kavli Institute for Theoretical Physics at Santa Barbara,
where this work was initiated during the Present and Future Neutrino Physics
Workshop. ANK is thankful to O. Smirnov and Aldo Ianii of Borexino
collaboration for the useful discussions and communication with them. The
financial support to ANK for this work has been provided by the Sun Yat-Sen
University under the Post-Doctoral Fellowship program.
\end{acknowledgments}

\end{document}